\newcommand{\farcs}{\mbox{\ensuremath{.\!\!^{\prime\prime}}}}

\documentclass[12pt,preprint]{aastex}

\slugcomment{Accepted for the ApJS}
\shorttitle{PTI Calibrator Catalog}
\shortauthors{van Belle et al.}

\begin{document}
\title{The Palomar Testbed Interferometer Calibrator Catalog}
\author{G. T. van Belle}
\affil{Michelson Science Center, California Institute of Technology,
      Pasadena, CA 91125} \email{gerard@ipac.caltech.edu}

\author{G. van Belle}
\affil{Department of Biostatistics, University of Washington,
Seattle, WA 98195-7232} \email{vanbelle@u.washington.edu}

\author{M. J. Creech-Eakman}
\affil{New Mexico Tech} \email{mce@kestrel.nmt.edu}

\author{J. Coyne}
\affil{University of Cambridge} \email{jc466@cam.ac.uk}

\author{A. F. Boden, R. L. Akeson, D. R. Ciardi, K. M. Rykoski, R. R. Thompson}
\affil{Michelson Science Center, California Institute of Technology,
      Pasadena, CA 91125} \email{bode, akeson, ciardi, kmr, thompson@ipac.caltech.edu}

\author{B. F. Lane}
\affil{Massachusetts Institute of Technology} \email{blane@MIT.EDU}

\and

\author{The PTI Collaboration}
\affil{Jet Propulsion Laboratory \& Michelson Science Center}

\begin{abstract}

The Palomar Testbed Interferometer (PTI) archive of observations
between 1998 and 2005 is examined for objects appropriate for
calibration of optical long-baseline interferometer observations -
stars that are predictably point-like and single. Approximately
1,400 nights of data on 1,800 objects were examined for this
investigation. We compare those observations to an intensively
studied object that is a suitable calibrator, HD217014, and
statistically compare each candidate calibrator to that object by
computing both a Mahalanobis distance and a Principal Component
Analysis.  Our hypothesis is that the frequency distribution of
visibility data associated with calibrator stars differs from
non-calibrator stars such as binary stars. Spectroscopic binaries
resolved by PTI, objects known to be unsuitable for calibrator use,
are similarly tested to establish detection limits of this approach.
From this investigation, we find more than 350 observed stars
suitable for use as calibrators (with an additional $\approx 140$
being rejected), corresponding to $\gtrsim 95\%$ sky coverage for
PTI. This approach is noteworthy in that it rigorously establishes
calibration sources through a traceable, empirical methodology,
leveraging the predictions of spectral energy distribution modeling
but also verifying it with the rich body of PTI's on-sky
observations.

\end{abstract}


\section{Introduction}

Visible and near-infrared interferometers are powerful tools for
measuring the minute angular sizes of nearby stars. However,
establishing absolute system responses in the presence of
atmospheric turbulence and instrument imperfections is a challenging
proposition that requires careful attention to detail when
constructing an observational approach.  Use of objects predicted to be point-like calibration
sources
is routinely employed in astronomical interferometry in the optical \citep{moz91,1998SPIE.3350..872B,2002A&A...393..183B, van05}.  However,
given the incomplete nature of our knowledge of these sources, it is not enough to merely
predict the expected fringe visibility from these objects - calibration sources need to be
rigorously evaluated for their actual observed interferometric visibility, and its appropriateness for
use in calibrating the instrument.
Herein we examine the body of calibration data taken
upon the sky from near-infrared, long-baseline interferometric
measurements taken with the Palomar Testbed Interferometer (PTI).
Similar efforts have begun for the Very Large Telescope Interferometer (VLTI) \citep{2005A&A...434.1201R}.

PTI is an 85 to 110 m H- and K-band (1.6 $\mu$m and 2.2 $\mu$m)
interferometer located at Palomar Observatory and is described in
detail in Colavita et al. (1999).  It has three 40-cm apertures used in pairwise
combination for measurement of stellar fringe visibility on sources that
range in angular size from 0.05 to 5.0 milliarcseconds, being able to resolve individual
sources $\theta \gtrsim 1.0$ mas in size.
PTI has been in nightly operation since 1997, with minimum downtime
for throughout the intervening years.
The data from PTI considered herein covers the range from the beginning of 1998 (when
the standardized data pipeline went into place) until the end of 2005 (when
the analysis of this manuscript was begun).  Over the 8 years of operation discussed
in this study, PTI was on the sky in its single-star visibility mode 1,390
nights\footnote{The remaining nights were spent on other instrument modes or
were lost to maintenance or weather.}, producing over
83,000 125 second stellar `scans' on 1,818 individual objects.
PTI has a minimum K-band fringe
spacing of $\approx 4.3$ mas at the sky position of the calibration
objects, making many of these stars readily resolvable.

As is standard practice in optical interferometry, a typical set of
observations with PTI involves observation of target objects of scientific interest, bracketed
by calibration objects.  The calibration objects serve the purpose of characterizing
the point-response, or system visibility ($V^2_{sys}$),  of the interferometer in
conjunction with the atmosphere, and as such, the interleaved
calibrator-target-calibrator observations are done on timescales expected to be shorter than
the seeing evolution time of the atmosphere (typically less than 15 minutes).
By measuring $V^2_{sys}$, a properly normalized measurement of the target star's
visibility may be obtained for meaningful astrophysical interpretation \citep{van05}.
These calibration objects are selected {\it a priori} to
be as close to point-like as is possible, having visibilities measured by PTI to be very nearly unity.
The primary limitation on selection of truly point-like calibrators is the sensitivity of the instrument,
coupled with considerations of dynamic range.
Towards the end of having nearly point-like calibrators,
calibrator lists are vetted for binary systems, which are excluded due to the
sub-unity system response measured from such systems.  The decrease of the measured visibility from
the desired characterization of $V^2_{sys}$
is due to the interferometer's ability to resolve out even milli-arcsecond separation binaries.

Unfortunately, incomplete knowledge of the true physical nature of the calibration objects can
result in binary systems not being properly excluded during the vetting process that selects
potential calibrators.  Given this reality of calibrator selection, a number of additional
steps are taken to ensure rigorously defensible characterization of $V^2_{sys}$.
First, given the uncertain nature of possible calibrators, employing multiple calibrators for any given
target star observation is also standard operating procedure for PTI observations.  Second,
these clusters of calibrators are observed on multiple nights, since chance geometry may
make a binary system readily apparent on one night but not on another.  Finally, these objects
may be compared to known `good' calibrators for appropriateness as calibration objects, which this
study will examine in detail.

\section{Approach}

To establish a properly normalize the
squared visibility\footnote{Squared visibility is referred to herein as simply, `visibility',
which is consistent with other articles in the literature.}
for a science target in the expected [0:1] range,
the system response is used to account for
the effect of instrumental and atmospheric imperfections and
normalize the visibility measured for that target, $V^2_{\textrm{\tiny meas}}$:
\begin{equation}\label{eqn_measSys}
V^2_{\textrm{\tiny norm}}(\textrm{target}) = { V^2_{\textrm{\tiny meas}}(\textrm{target}) \over V^2_{\textrm{\tiny sys}} }
\end{equation}
The system visibility, $V^2_{\textrm{\tiny sys}}$, is established from measurements of point-like calibration
sources.  However, due to the extreme resolution of interferometric instruments, event nominally `unresolved'
objects need to have their measured $V^2_{\textrm{\tiny meas}}$ corrected for partial resolution:
\begin{equation}\label{eqn_measExp}
V^2_{\textrm{\tiny sys}} = { V^2_{\textrm{\tiny meas}}(\textrm{calibrator}) \over V^2_{\textrm{\tiny pred}}(\textrm{calibrator}) }
\end{equation}
Interleaved observations of targets and calibrators are done
on a time scale expected
to be less than the atmospheric seeing evolution time, which at the Palomar
site is typically 30 minutes or longer.  PTI's automated star queue duty cycle is 4 to 8 minutes per star,
so calibrator-target-calibrator interleaving is comfortably accomplished
with sufficient cadence to satisfy this requirement.

Assuming that a uniform disk brightness profile appropriately characterizes the stellar disk
of the calibration source, the predicted visibility $V^2_{\textrm{\tiny pred}}$
may be derived from some expectation of the star's angular size, $\theta$, and
the known system configuration parameters of baseline $B$ and wavelength, $\lambda$:
\begin{equation}\label{eqn_UDdisk}
 V^2_{\textrm{\tiny pred}}
     = \left[{2 J_1({\pi \theta B / \lambda }) \over {\pi \theta B / \lambda }}\right]^2
\end{equation}
In estimating $V^2_{\textrm{\tiny pred}}$, and by extension $V^2_{\textrm{\tiny sys}}$,
it is important to properly quantify errors in $B$, $\lambda$, and $\theta$,
and propagate them through the Bessel function.  The need for `point-like' calibrators
is prompted by potential unknown biases present in the {\it a priori} estimation
of the calibrator angular size, $\theta$, and to avoid non-linear effects found
in the Bessel function.
Proper error propagation is observed throughout all of these steps, including
propagation of uncertainties in angular size estimate, $\sigma_\theta$,
into $V^2_{\textrm{\tiny pred}}$ and measurement error, $\sigma_{V^2}$, for both $V^2_{\textrm{\tiny meas}}(\textrm{target})$
and $V^2_{\textrm{\tiny meas}}(\textrm{calibrator})$.
The particulars and implications
of this requirement are examined in much greater detail in \citet{van05}.
In practice, for calibrators previously uncharacterized by PTI, 2 or more calibration sources are used in tandem
to all for cross-calibration and elimination of bad calibrators.

For this study, the full list of stars observed by PTI from 1998 to 2005 was collected from
the PTI archive\footnote{Online at the Michelson Science Center, http://msc.caltech.edu.}.
A set of calibrator selection criteria was established (\S \ref{sec_CalSelCritera}), and
for
each of the objects in the archive that qualified as a potential calibrator,
 a spectral energy distribution (SED) fit was used to refine its predicted
angular size (\S \ref{sec_SED_fitting}).  Objects which were small enough for
use as calibrators were kept as possible calibrators.  The predicted angular size
could then be used to account for the partial resolution by the interferometer of
the object's angular size in attempting to normalize the object's
raw interferometer data (\S \ref{sec_NormPTIObsvns}).
The data obtained from the normalization step were then examined with Mahalanobis distance
and Principal Component Analysis tests for departures from
point-source response and evidence of hidden binarity (\S \ref{sec_StatTests}).
Finally, we will examine the likelihood that stars observed by PTI that appear to be solid
calibrators are actually undetected binaries (\S \ref{sec_UndetectedBinaries}).

\section{Calibrator Selection Criteria}\label{sec_CalSelCritera}
In order to maximize the likelihood that any random star observable by PTI is appropriate
for use as a calibrator, a number of criteria have been developed over the years of operation
of the instrument. Since each principal investigator selected his or her own calibration sources
for their projects, it is unclear the exact criteria that went into
the selection of both targets and calibrators found in the complete PTI archive.  However, since we
are evaluating the sources after the fact, we may vet the observed list of potential calibrators
using a single homogenous set of criteria:
\begin{itemize}
\item Potential calibration sources must be bright enough to be tracked by PTI in the appropriate available dynamic range.
This translates roughly to a $V$ band magnitude
of $\sim 10.0$ or brighter (for tip-tilt tracking), and a $K$ band magnitude of $\sim 5.0$
or brighter (for fringe tracking).
\item These sources need to be appropriately `point-like' in size, which conservatively for PTI is less
than $\lesssim 1.0$ mas in size, as discussed extensively in \citet{van05},
with final predicted size from detailed SED fitting (see \S \ref{sec_SED_fitting}).  This corresponds to a
measured visibility $V^2\gtrsim 0.90$ that serves well to characterize the system visibility
(interferometer plus atmosphere) and is statistically tolerant to errors and even biases in the
{\it a priori}
estimation of the calibrator size.
\item Finally, for system visibility characterization, these stars must be expected to
exhibit constant measured visibilities.
As such, they should not be known to be a binary system, or suspected to be one, according to
the astrometric references in the Hipparcos database \citep{1997A&A...323L..49P}.
The Hipparcos H59 multiplicity flag is of particular utility here, not only
calling out those objects for which a full astrometric solution has been obtained, but those
which have unexplained degrees of astrometric variability.
\end{itemize}
The 499 potential sources in the PTI database which satisfy these criteria
are found in Table \ref{tab_SED_photom}, the contents of which will be discussed in the next section.
The PTI data available for these sources will then be evaluated in \S \ref{sec_StatTests}
for evidence of binarity or departures in {\it a priori} size expectations that
make them unsuitable for use as calibrators.

\section{Spectral Energy Distribution Fitting}\label{sec_SED_fitting}

For each of the potential calibrator stars observed in this investigation, a
spectral energy distribution (SED) fit was performed.  This fit was
accomplished using photometry available in the literature as the
input values, with template spectra appropriate for the spectral
types indicated for the stars in question. The template spectra,
from \citet{pic98}, were adjusted by the fitting routine to account
for overall flux level, wavelength-dependent reddening, and expected
angular size.  Reddening corrections were based upon the empirical
reddening determination described by \citet{1989ApJ...345..245C}, which differs
little from van de Hulst's theoretical reddening curve number 15
\citep{joh68,dyc96}. Both narrowband and wideband photometry in the
0.3 $\mu$m to 30 $\mu$m were used as available, including Johnson
$UBV$ (see, for example, \citet{1963AJ.....68..483E,1972ApJ...175..787E,1971A&A....12..442M}),
Stromgren $ubvy\beta$
\citep{1976HelR....1.....P}, 2Mass $JHK_s$ \citep{2003yCat.2246.....C}, Geneva \citep{1976A&AS...26..275R},
Vilnius $UPXYZS$ \citep{1972VilOB..34....3Z}, and $WBVR$ \citep{1991TrSht..63....4K}; flux calibrations
were based upon the values given in \citet{cox00}.  For each star, the
primary references for the photometric data are given in Table \ref{tab_SED_photom}.


\begin{deluxetable}{r|cccccc|p{1.5in}|p{0.5in}|p{1.5in}}
\rotate
\tabletypesize{\scriptsize}
\tablecolumns{10}
\tablewidth{0pc}
\tablecaption{Sources of photometry for stars tested as PTI calibrators.\label{tab_SED_photom}}
\tablehead{
\colhead{HD} & \colhead{WBVR}   & \colhead{Geneva}    & \colhead{DDO} &
\colhead{Vilnius} & \colhead{2MASS}   & \colhead{IRAS}    & \colhead{Johnson}   & \colhead{Stromgren}    & \colhead{JP11}}
\startdata

        71 &          Y &          - &          - &          - &          - &          Y &            &            &            \\

       166 &          Y &          Y &          Y &          - &          Y &          Y & 90, 18, 19, 92, 243 &            & 270, 262, 230 \\

       167 &          Y &          - &          Y &          - &          Y &          Y &            &            &            \\

       905 &          Y &          Y &          - &          Y &          - &          Y &        107 &            &    270, 87 \\

      1083 &          Y &          - &          - &          - &          Y &          - &            &            &   270, 146 \\

      1279 &          Y &          Y &          - &          Y &          Y &          - &         28 &            & 270, 146, 161 \\

      1404 &          Y &          Y &          - &          - &          Y &          Y & 90, 6, 18, 88, 90, 243 &         90 & 270, 129, 87, 246, 260 \\

      1406 &          Y &          Y &          Y &          - &          Y &          Y &            &            &            \\
\enddata
\end{deluxetable}

\begin{figure}
\includegraphics[scale=.66,angle=270]{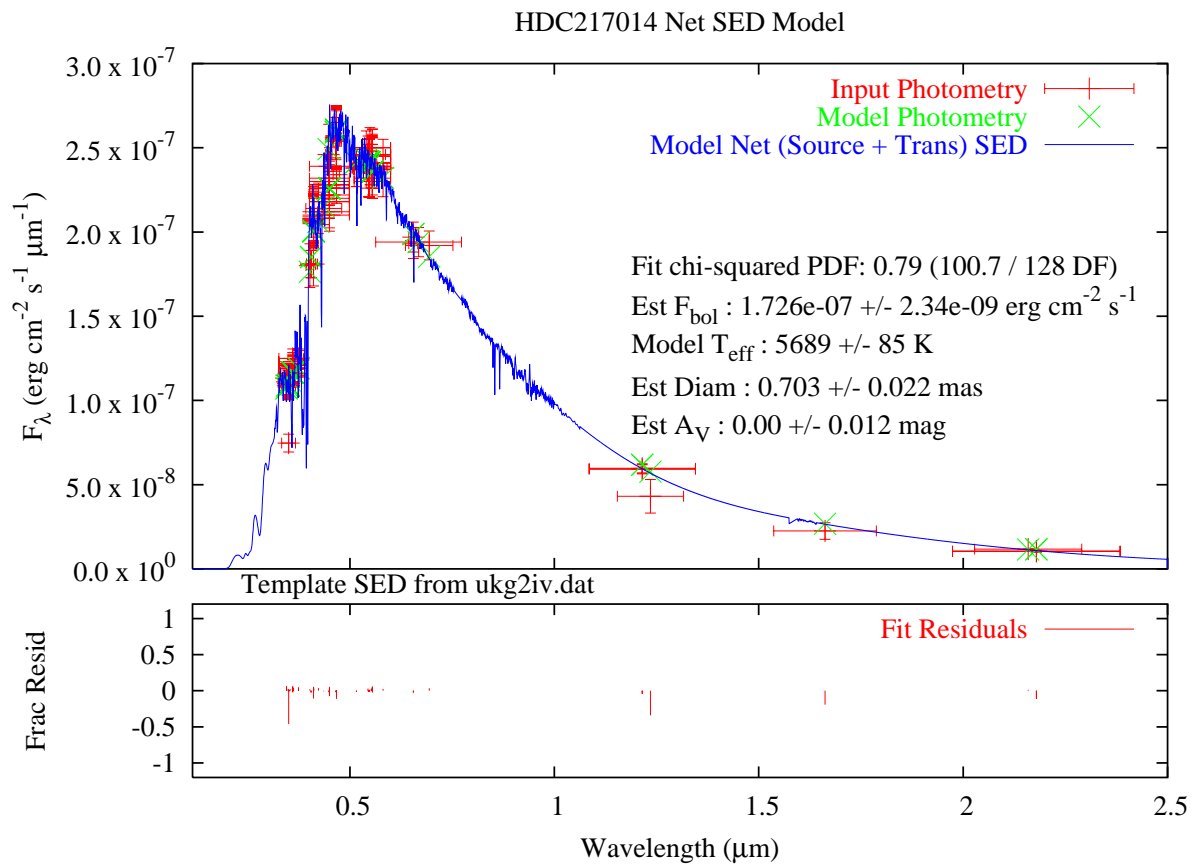}
\caption{\label{fig_HD217014} Spectral energy distribution fitting for HD 217014,
as discussed in \S \ref{sec_51Peg}, with a G2 IV spectral template \citep{pic98} being
fit to the wide- and narrow-band photometry available for the star.
Vertical bars are errors associated with
the photometric data; horizontal bars represent the bandwidth of each
photometric data point.}
\end{figure}

Starting with a reference spectral type and luminosity class as cited by SIMBAD,
template spectra were fit to the photometric data.  Templates in adjacent
locations in spectral type and luminosity class were also
tested for best fit, with the fit with best $\chi^2$ being selected in the end
for use in this study.  For example, a star indicated by SIMBAD to be a G2IV would have its
photometry fit to the 9 templates of spectral type G1, G2 and G3, and for luminosity classes
V, IV, and III.
From the best SED fit, estimates were obtained for each star for their bolometric
flux ($F_{BOL}$), angular size ($\theta_{EST}$), and reddening ($A_V$);
effective temperature was fixed for each of the \citet{pic98} library spectra.
The results of
the fitting are given in Table \ref{tab_SED_fits}, and an example
SED fitting plot is given in Figure \ref{fig_HD217014}.

Also given in Table 2 are estimates of luminosity for each of these
stars.  The first estimate was derived from the bolometric flux from
the fitting, combined with a Hipparcos distance \citep{1997A&A...323L..49P}; a
small number of these objects ($N=11$) did not have such data available
and a distance was estimated from comparison of the apparent visible
brightness $m_V$ to the visible $M_V$ brightness expected from the
best-fit spectral type \citep{cox00}, and the SED fit reddening.  The second estimate was
established from the luminosity expected from the best-fit spectral type.
The agreement between these two disparate estimates
provided us with additional confidence in our fits, particularly
from the standpoint of selection of proper luminosity class for the SED fit.



\begin{deluxetable}{ccccccccc}
\tablecolumns{9}
\tabletypesize{\scriptsize}
\tablewidth{0pc}
\tablecaption{\label{tab_SED_fits}Summary of spectral energy distribution fits for potential PTI calibrator stars, including
total number of photometric points used, $N_{PHOT}$,
reduced chi-squared $\chi^2_\nu$, reddening $A_V$, bolometric flux at the source $F_{BOL}$,
luminosity estimate from bolometric flux $L_{FBOL}$ and from spectral type $L_{SP}$, and estimated angular size $\theta_{EST}$.}
\tablehead{
\colhead{HD} & \colhead{Spectral} & \colhead{$N_{PHOT}$} & \colhead{$\chi^2_\nu$} & \colhead{$A_V$} & \colhead{$F_{BOL}$} & \colhead{$\log L_{FBOL}$} & \colhead{$\log L_{SP}$}& \colhead{$\theta_{EST}$} \\
\colhead{} & \colhead{Type} & \colhead{} & \colhead{} & \colhead{(mag)} & \colhead{($10^8$ erg cm$^{-2}$ s$^{-1}$)} &\colhead{$(L_\odot)$} &\colhead{$(L_\odot)$} & \colhead{(mas)}
 } \startdata


        71 &      K1III &          3 &       4.39 & $ 0.38 \pm 0.07 $ & $ 7.45 \pm 0.17 $ & $ 2.11 \pm 0.11 $ & $ 1.88 \pm 0.50 $ & $ 0.690 \pm 0.036 $ \\

       166 &        G8V &         41 &       0.35 & $ 0.02 \pm 0.02 $ & $ 10.46 \pm 0.08 $ & $ -0.21 \pm 0.04 $ & $ -0.14 \pm 0.25 $ & $ 0.626 \pm 0.013 $ \\

       167 &      G8III &          8 &       0.76 & $ 0.13 \pm 0.03 $ & $ 7.34 \pm 0.09 $ & $ 1.94 \pm 0.50 $ & $ 1.74 \pm 0.50 $ & $ 0.594 \pm 0.036 $ \\

       905 &        F0V &         42 &       0.85 & $ 0.09 \pm 0.01 $ & $ 14.40 \pm 0.13 $ & $ 0.74 \pm 0.02 $ & $ 0.85 \pm 0.25 $ & $ 0.400 \pm 0.010 $ \\

      1083 &       A0IV &         16 &       0.44 & $ 0.03 \pm 0.02 $ & $ 8.54 \pm 0.14 $ & $ 1.63 \pm 0.34 $ & $ 2.46 \pm 0.33 $ & $ 0.167 \pm 0.010 $ \\

      1279 &      B5III &         37 &       0.95 & $ 0.30 \pm 0.01 $ & $ 34.98 \pm 0.25 $ & $ 2.97 \pm 0.62 $ & $ 3.86 \pm 0.50 $ & $ 0.153 \pm 0.025 $ \\

      1404 &        A2V &         55 &       1.00 & $ 0.08 \pm 0.03 $ & $ 44.12 \pm 0.33 $ & $ 1.41 \pm 0.10 $ & $ 1.46 \pm 0.25 $ & $ 0.465 \pm 0.023 $ \\

      1406 &      K2III &         15 &       1.30 & $ 0.23 \pm 0.02 $ & $ 6.87 \pm 0.08 $ & $ 1.74 \pm 0.46 $ & $ 1.94 \pm 0.50 $ & $ 0.728 \pm 0.037 $ \\
\enddata
\tablenotetext{a}{No Hipparcos distance available; distance estimated from comparison of apparent $m_V$ with absolute $M_V$ expected from the spectra type.}
\end{deluxetable}



For our calibration
sources, the {\it a priori} estimate of their angular size $\theta_{EST}$ is
necessary to account for residual resolution that may be afforded by
an interferometer's extraordinarily long baselines.  With an
expected limb darkened size of $\theta_{EST} \leq 1.00$ mas from the
SED fit, calibrators have predicted $V^2_{\textrm{\tiny pred}}$ values of $>92$\% for a
85-m baseline used at 2.2 $\mu$m, and $>86$\% for a 110-m baseline.  We shall consider this size
effectively identical to its uniform disk size, since for
most of our potential calibration sources,
their effective temperatures are in excess of $\sim 5000$K,
and the difference between the uniform disk and limb darkened sizes is at the few percent level
\citep{dav00,cla03b}, which is far less than our size estimate
error. Ideally, a calibration source would be sufficiently
point-like that its measured visibility $V^2_{\textrm{\tiny meas}}$ would be indistinguishable from unity, but
unfortunately the current system sensitivity does not afford that
option.  The uncertainty in the calibrator visibility represents one
of the fundamental limitations of the system visibility accuracy \citep{van05}.
However, our selection of calibrators has been carefully made to be sufficiently small ($\leq 1.00$ mas)
in diameter such that there are no concerns about a varying system
calibration due to unaccounted-for calibrator surface morphology,
and such that a $\leq 5\%$ uncertainty in angular size will translate to
less than a $\leq 1\%$ uncertainty in its predicted visibility $V^2_{\textrm{\tiny pred}}$ for PTI.

\section{Obtaining Normalized PTI Observations}\label{sec_NormPTIObsvns}

PTI generates its $V^2_{\textrm{\tiny meas}}$ observables through pairwise combination
of the starlight collected by two siderostats through a
50/50 beam combiner optic; the combined starlight beams that result are
fed into a NICMOS3 detector dewar.  The PTI beamtrain
from the siderostats to the beam combiner
includes
pathlength compensation that equalizes the path for each telescope
from the star to the beam splitter to $\sim 20$ nm.  Temporally dithering the
pathlength compensation within one atmospheric coherence time
a distance of one wavelength about this `white light'
fringe position results in maximum constructive and destructive interference between
the two starlight beams.  Measurement of this temporally modulated
photometric signal leads to a characterization of the $V^2_{\textrm{\tiny meas}}$.
Discussion of the PTI fringe detection and tracking particulars
is given in significantly greater detail in \citet{col99}.

The $V^2_{\textrm{\tiny meas}}$ observables used in any typical PTI study are
the synthetic wideband $V^2_{\textrm{\tiny meas}}$ values, given by an incoherent
signal-to-noise (SNR)
weighted average of the individual
5 narrowband channel $V^2_{\textrm{\tiny meas}}(\lambda)$ values in the PTI
spectrometer \citep{col99b}.  In a similar fashion,
incoherent
SNR-weighted average bandpasses $\lambda$ were determined from the raw data.
The PTI
H and K wavebands are excellent matches to the CIT photometric
system \citep{eli82,eli83}. Separate
calibrations and fits to the narrowband and synthetic wideband
$V^2_{\textrm{\tiny meas}}$ data sets yield statistically consistent results, with the
synthetic wideband data exhibiting superior SNR.
Since there is no need in this study for independent narrowband values,
we will consequently consider only the synthetic
wideband data.

The stars examined in this study were observed by PTI
at 1.6 and 2.2~$\mu$m on 1,390 observing nights between 1998 Jan
1 and 2005 Dec 31.  For most of the nights, PTI's NS 110m baseline (37.1 m E, 103.3 m N, -3.3 m Z)
was utilized; the NW 86m baseline (-81.7 m E, -28.2 m N, 3.1 m Z) was used the second most, with the SW 87 m baseline
(-44.6 m E, 75.1 m N, -0.2 m Z)
being used least.
The specific baselines used, in terms of number of nights and number of scans, is
given in Table \ref{tab_PTI Data} for each star; a summary of the overall observations is given
in Table \ref{tab_obs_sum}.
The potential calibration objects were
observed multiple times during each of these nights, and each
observation, or scan, was approximately 125 s long. For each scan
we computed a mean $V^2_{\textrm{\tiny meas}}$ value from the scan data, and the error
in the $V^2_{\textrm{\tiny meas}}$ estimate from the rms internal scatter \citep{col99, col99b}.



\begin{deluxetable}{rrrrrrrrrrrr}
\tablecolumns{5}
\tabletypesize{\scriptsize}
\tablewidth{0pc}
\tablecaption{\label{tab_PTI Data}Summary of PTI observations for each potential calibrator,
including nights observed and number of calibrated scans, both overall and by baseline.}
\tablehead{
\colhead{}   & \colhead{Raw}         & \multicolumn{2}{c}{\underline{All Baselines}}           & \multicolumn{2}{c}{\underline{North-South}}            & \multicolumn{2}{c}{\underline{North-West}}             & \multicolumn{2}{c}{\underline{South-West}}                   \\
\colhead{}   & \colhead{Nights}   & \colhead{Calibrated} & \colhead{Calibrated} & \colhead{Calibrated}& \colhead{Calibrated} & \colhead{Calibrated} & \colhead{Calibrated}& \colhead{Calibrated}   & \colhead{Calibrated} \\
\colhead{HD} & \colhead{Observed} & \colhead{Nights}     & \colhead{Scans}      & \colhead{Nights}    & \colhead{Scans}      & \colhead{Nights}     & \colhead{Scans}     & \colhead{Nights}       & \colhead{Scans}
}
\startdata


        71 &                2  &                2  &                5  &              -    &              -    &                2  &                5  &              -    &              -    \\

       166 &              67  &              67  &            270  &              53  &            217  &              12  &              49  &                2  &                4  \\

       167 &                1  &                1  &                2  &                1  &                2  &              -    &              -    &              -    &              -    \\

       905 &                4  &                4  &              11  &                3  &                7  &                1  &                4  &              -    &              -    \\

      1083 &                1  &                1  &                2  &                1  &                2  &              -    &              -    &              -    &              -    \\

      1279 &                9  &                9  &              24  &                9  &              24  &              -    &              -    &              -    &              -    \\

      1404 &            107  &            104  &            367  &              71  &            243  &              29  &            108  &                4  &              16  \\

      1406 &                2  &                2  &              12  &                2  &              12  &              -    &              -    &              -    &              -    \\
\enddata
\end{deluxetable}





\begin{deluxetable}{lcrrr}
\tablecolumns{5}
\tabletypesize{\scriptsize}
\tablewidth{0pc}
\tablecaption{\label{tab_obs_sum}Summary of observations.}
\tablehead{
\colhead{} & \colhead{Number}                 & \multicolumn{3}{c}{\underline{Proportion}\tablenotemark{a}}      \\
\colhead{Baselines}   & \colhead{of Stars}    & \colhead{NS}    & \colhead{NW}   & \colhead{SW}
 } \startdata


   NS only &        148 &          1.00  &                -  &                -  \\

   NW only &         88 &                -  &          1.00  &                -  \\

   SW only &          1 &                -  &                -  &          1.00  \\

    NS, NW &        135 &          0.63  &          0.37  &                -  \\

    NS, SW &          5 &          0.66  &                -  &          0.34  \\

    NW, SW &          7 &                -  &          0.52  &          0.48  \\

NS, NW, SW &        105 &          0.60  &          0.32  &          0.08  \\

None (uncalibrated) &         10 &                -  &                -  &                -  \\

\enddata
\tablenotetext{a}{Distribution, on average, of the observations for each of the 3 PTI baselines. }
\end{deluxetable}



\section{Statistical Tests for the Observational Data}\label{sec_StatTests}

\subsection{51 Peg: The PTI `Gold Standard' Calibrator}\label{sec_51Peg}

51 Peg (HD 217014, HP 113357, HR 8729) is well-known throughout the astronomical community
as the first solar-like star for which an extrasolar planet
has been detected \citep{1995Natur.378..355M, 1997ApJ...481..926M}.
When the sinusoidal radial velocity signature was detected in 1995,
the $\sin i$ ambiguity in the mass term led some to suspect that
perhaps what was being observed was a  binary star system in a face-on orientation, rather
than a star-planet pairing.

Interferometric observations have the potential to
detect such binary star systems through spatially resolving and
directly detecting the individual components.
As such, when PTI operations began in 1996, a vigorous campaign of 51 Peg
observations was begun to attempt to detect any putative stellar
companion of the primary star.  Despite some ambiguous initial indications
associated with a new instrument learning curve,
in the end a rigorous evaluation of the PTI data excluded the
presence of a companion brighter than $\Delta K<4.27$ (corresponding
to a brightness ratio $r(K)>0.020$) at the 99\% confidence level,
corresponding to an upper mass limit
on the 51 Peg secondary of 0.22 $M_\odot$ \citep{bod98}.
Given the unusually detailed evaluation of the 18 nights available (at that time) of PTI data on
51 Peg, it now constitutes a `gold standard' calibration source,
as far as being an object whose nature as a system possessing
a single star has been established to the limits of PTI's
detection ability.  For this particular investigation, we
may leverage that detailed investigation of 51 Peg to use it as a reference calibration object,
to which other potential calibrators may be compared.

\subsection{Known Bad Calibrators: PTI Binaries}\label{sec_PTIBinaries}

In contrast to 51 Peg, we may also select well-studied PTI sources
as known bad calibrators - namely, the binary stars that PTI has
observed over the years. The full roster of published PTI
binaries, along with 51 Peg, are summarized in Table
\ref{tab_PTI_binaries}.  Also included on this list is an unpublished
binary star, HD178449, which with a primary-secondary brightness ratio
of $\Delta K \sim 3.5-3.8$ represents the `worst case' binary for our various
methods to attempt to detect; as will be discussed in more
detail on \S \ref{sec_UndetectedBinaries},
this level of $\Delta K$ is consistent with
the median value of $V^2_{\textrm{\tiny meas}} \sim 0.93-0.95$
in Equation \ref{eqn_binaries}.

These known binary stars are of great utility to the evaluation of
our unknown potential calibrator objects.  By subjecting these known binary stars to our statistical
tests for appropriateness of a calibration star, we may test the sensitivity
of those tests to binaries of varying brightness ratio and separation.



\begin{deluxetable}{ccccl}
\tablecolumns{5}
\tablewidth{0pc}
\tablecaption{Summary of spectroscopic binary stars spatially resolved by PTI.\label{tab_PTI_binaries}}
\tablehead{
\colhead{Star} & \colhead{HD} & \colhead{a (mas)} & \colhead{$\Delta K$} & \colhead{Reference}
 } \startdata
    64 Psc &       4676 &       6.50 &       0.11 & \citet{bod99b} \\

           &       6118 &       5.56 &       0.40 & \citet{kon04} \\

    TZ Tri &      13480 &       2.10 &       1.75 & \citet{kor98}\tablenotemark{a} \\

     Atlas &      23850 &      12.94 &       1.86 & \citet{pan04} \\

           &      27483 &       1.26 &       0.00 & \citet{kon04} \\

   omi Leo &      83808 &       4.46 &       1.49 & \citet{hum01} \\

    12 Boo &     123999 &       3.39 &       0.61 & \citet{bod00, bod05} \\

           &     178449 &      n/a &       $\sim 3.5-3.8$ & Unpublished\\

           &     195987 &      15.38 &       1.06 & \citet{tor02} \\

  iota Peg &     210027 &      10.33 &       1.61 & \citet{bod99a} \\

    BY Dra &     234677 &       4.40 &       0.57 & \citet{bod01} \\
\hline
    51 Peg &     217014 &        n/a &  $> 4.27$ & \citet{bod98} \\

\enddata
\tablenotetext{a}{See discussion in \S \ref{sec_PTIBinaries}.}
\end{deluxetable}



\subsection{Moment Values for 51 Peg}\label{sec_51PegMoments}

For detailed statistical testing of the visibility data on each of
our potential calibrator stars found in Table \ref{tab_SED_fits}, we sought to
compare the ensemble of 51 Peg $V^2_{\textrm{\tiny norm}}$ points with the various sets
associated each potential calibrator star first by means of a
Mahalanobis distance comparison, and second through a Principal
Component Analysis. The variable of interest is an individual normalized PTI
observation, $V^2_{\textrm{\tiny norm},i}$, with associated error term
$\sigma_{V^2,i}$.  For each star we have a number of observations
of $V^2_{\textrm{\tiny norm}}$ that form a frequency
distribution. Our hypothesis is that the frequency distribution associated with
calibrator stars differs from non-calibrator stars such as binary stars. The challenge
is to characterize the frequency distribution. A very old technique is that of using
Pearson curves which are characterized by the first four moments of a distribution.
A function of the raw moments of the frequency distribution of $V^2_{\textrm{\tiny norm}}$ for each
star forms the input for our comparison with the first four moments of 51 Peg $V^2_{\textrm{\tiny norm}}$ data.

The observations were weighted by
$w^*_i=1/(\sigma_{V^2,i})^2$. If $W=\sum_i w_i $ for specific $k$
star, and $w_i=w^*_i/W$, then the weighted mean, $m_1$, of
$V^2_{\textrm{\tiny norm},i}$ is $m_{1k}=\sum_i w_iV^2_{\textrm{\tiny norm},i}$. The weighted
second, third and fourth moments for this star are
\begin{equation}
m_{jk}=\left[\sum_i w_i(V^2_{\textrm{\tiny norm},i}-m_1)^j\right]^{1/j}
\end{equation}
for $j=2,3,4$. For $m_{3k}$ we retained the sign associated
with the third moment. The rescaled moments have a natural
interpretation; for example, $m_{2k}$, is the standard deviation of
the $V^2_{\textrm{\tiny norm}}$ values for star $k$.

These 
moments $m_{1k}, m_{2k}, m_{3k}, m_{4k}$ form the basis
for the calculation of the
Mahalanobis distance and the Principal Component Analysis distance from 51 Peg,
our `gold
standard'. The four 
moment values for 51 Peg were calculated to be,
\begin{eqnarray*}
m_{1,\textrm{\tiny 51 Peg}}&=&1.006659\\
m_{2,\textrm{\tiny 51 Peg}}&=&0.041568\\
m_{3,\textrm{\tiny 51 Peg}}&=&0.032527\\
m_{4,\textrm{\tiny 51 Peg}}&=&0.0919526
\end{eqnarray*}

The moments can be used to calculate $a_3=skewness$ and
$a_4=kurtosis$ statistics. For a Gaussian distribution the values
are $a_3=0$ and $a_4=3$. For the $V^2_{\textrm{\tiny norm}}$ values from 51 Peg these values are,
\begin{eqnarray*}
a_3=&{m_{3k} / {m_{2k}}^{3/2}}&= 1.32\\
a_4=&{m_{4k} / {m_{2k}}^{2}}&= 12.1
\end{eqnarray*}
Thus these data suggest that the 51 Peg $V^2_{\textrm{\tiny norm}}$
data are skewed to the right and leptokurtic.

\begin{figure}
\begin{center}
\includegraphics[scale=1.2,angle=0]{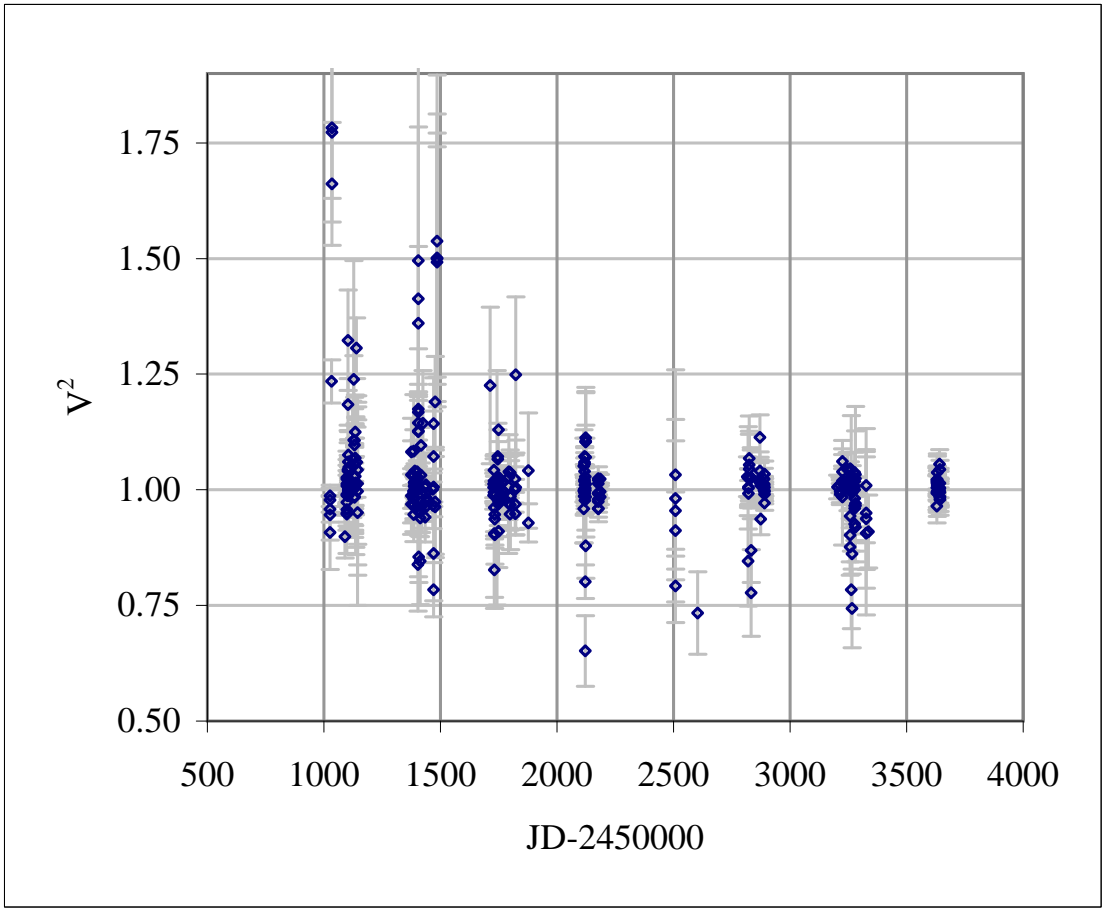}
\end{center}
\caption{\label{fig_51PegV2norm} Normalized visibility measurements ($V^2_{\textrm{\tiny norm}}$)
for 51 Peg,
as discussed in \S \ref{sec_51PegMoments}.}
\end{figure}

The skewness of the 51 Peg data is a result of there being a few, but notable, outlying
data points with $V^2_{\textrm{\tiny norm}} \gg 1$, as seen in Figure \ref{fig_51PegV2norm}.  These data points are all associated with
periods of poor system visibility when the system visibility was low, typically
$V^2_{\textrm{\tiny sys}} < 0.60$, which is usually a result of poor seeing; of the 340 data points in the 51 Peg dataset, 20
were in the low system visibility category.   Up
until this point in the analysis, all of the
data from 51 Peg contributed to the statistical analysis, in order to examine all
nights in a homogenous manner, regardless of atmospheric conditions.
However, if we use that parameter as a guide and exclude data points with
$V^2_{\textrm{\tiny sys}} < 0.60$ from that analysis, we find $a_3 = -0.048$ with
$a_4 = 6.13$ --- a negligible amount of skewness.

\begin{figure}
\begin{center}
\includegraphics[scale=1.2,angle=0]{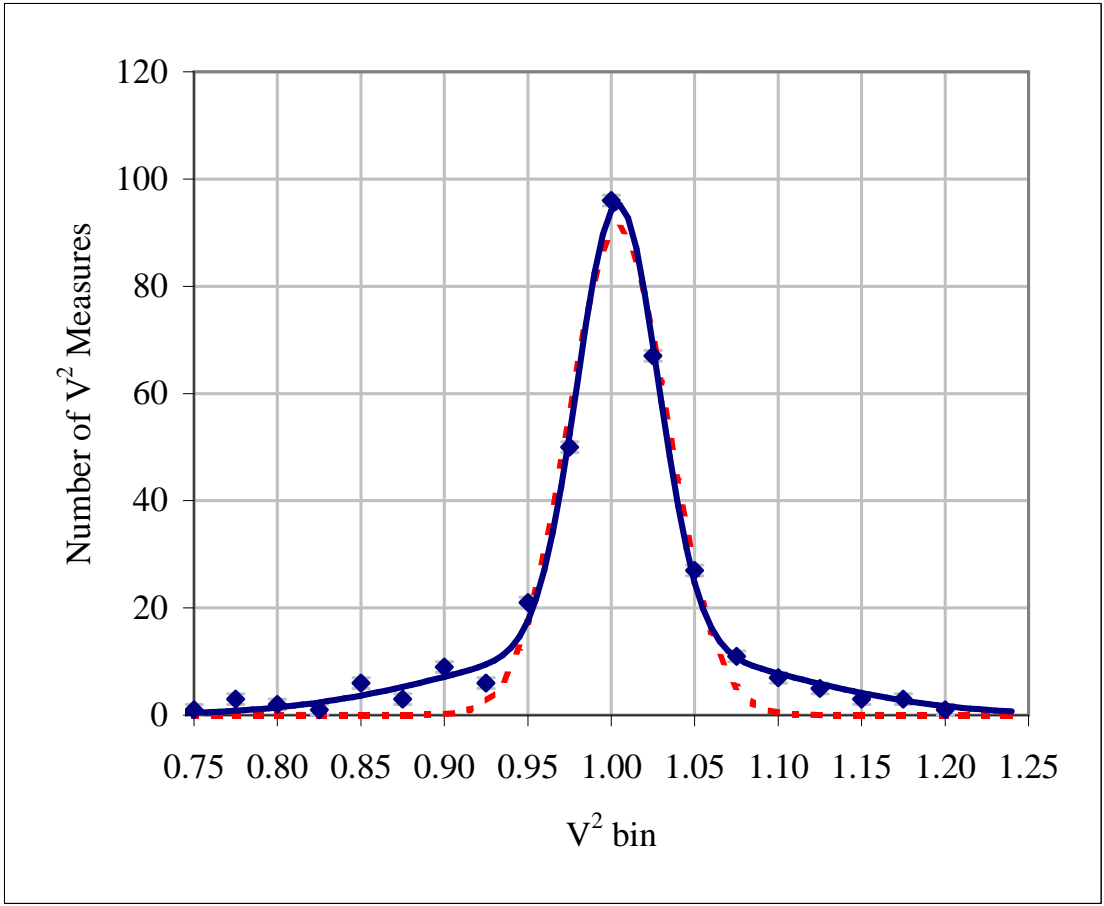}
\end{center}
\caption{\label{fig_51PegV2hist} Histogram of the 51 Peg normalized visibility measurements
($V^2_{\textrm{\tiny norm}}$)
as discussed in \S \ref{sec_51PegMoments}, in bins of $\Delta V^2=0.05$.  The dotted line is a single Gaussian fit, with
amplitude of 90.9 and width 0.0297, and the solid line is a two-Gaussian fit with amplitude/width
of 82.8/0.0242 and 12.5/0.098, respectively.  The $\chi^2$ per degree of freedom for the former is 5.44,
with $\chi^2/$DOF for the latter being 1.32.}
\end{figure}

The leptokurtic nature of the 51 Peg $V^2_{\textrm{\tiny norm}}$
data is seen with both the full ensemble and the $V^2_{\textrm{\tiny sys}} > 0.60$ subset ---
in terms of shape, a leptokurtic distribution has a more acute peak around the mean
and longer tails.  A histogram of the full 51 Peg $V^2_{\textrm{\tiny norm}}$ dataset is seen in
Figure \ref{fig_51PegV2hist}, in bins of $\Delta V^2=0.05$. Two fits were compared to this
binned data: the first was single Gaussian, which resulted in a mean of $\overline{V}^2=1.004$
and a width of 0.030 on an amplitude of 90.9, and a reduced $\chi^2$ per degree of freedom of
$\chi^2/$DOF=5.44; the second
was two Gaussians, with an identical mean but widths of 0.024 and 0.098 on amplitdues of 82.8 and 12.5,
respectively, with $\chi^2/$DOF=1.32.  The second fit suggests that two sources of
noise are present in the data.  Restricting these fits to the $V^2_{\textrm{\tiny sys}} > 0.60$ subset,
we find that the single Gaussian fit has a width of 0.029 with a $\chi^2/$DOF=5.02, and
the two-Gaussian fit has widths of 0.080 and 0.023 with $\chi^2/$DOF=1.22; no notable change for either
fit resulted in the amplitudes or $\overline{V}^2$ values.

These latter fits suggest that
for those times when the interferometer is performing best (resulting in better
$V^2_{\textrm{\tiny sys}}$ values) the wider `pedestal' Gaussian of the two-Gaussian fit
reduces in width.  Since the instrument is operated in a fairly uniform manner and
reports lowered $V^2_{\textrm{\tiny sys}}$ values generally during
periods of particularly poor atmospheric seeing,
our suspicion is that the error pedestal is associated
with those occasional periods of poor seeing.  Such observations
do not get properly normalized due to limitations of the calibration process for coping with those
occasional rapid fluctuations in seeing, but are identifiable in the data set
in part due to their low $V^2_{\textrm{\tiny sys}}$ values.  Further examination of the 51 Peg dataset
can exclude other measurement outliers (and the spread on $V^2_{\textrm{\tiny norm}}$ will
then approach the $\sim 1.5$\% value seen in \citep{bod98}) but this is time-consuming and
potentially quite subjective.  As such, we will maintain the integrity of the normalized
visibilities being
examined for all sources and not cut for $V^2_{\textrm{\tiny sys}}$
in the following sections.

\subsection{Mahalanobis and Principal Component Analysis Distances}
\begin{figure}
\includegraphics[scale=1.0]{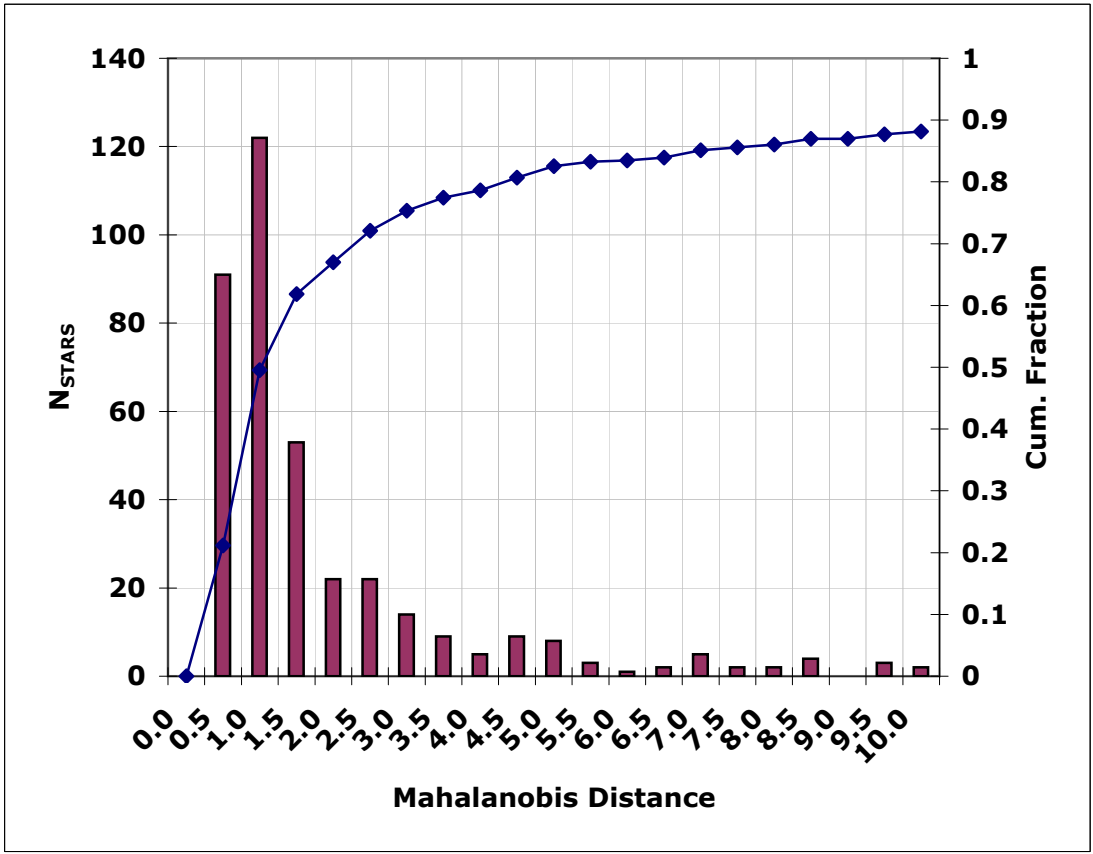}
\caption{\label{fig_Maha} Histogram of Mahalanobis distances from HD217014 for potential calibrator stars.}
\end{figure}

The Mahalanobis distance (MD) is a multivariate generalization of
one-dimensional Euclidean distance
\citep{1989AcA....39...85B, 1999MNRAS.303..284R}.
Given $N$ stars characterized by $M$ variables
(coordinates) $m_{jk}$, the sample distance of a particular star,
$k$, from the centroid of the distances is,
\begin{equation}
MD_k =
\sum_{j=1}^M{\left(\frac{m_{jk}-\overline{m_j}}{s_j}\right)^2}
\end{equation}
where $\overline{m_j}$ is the mean for the $j$th variable, and $s_j$
is its standard deviation. Alternatively, the MD of one star from a
specific star $k^*$, in this case 51 Peg say, is given by,
\begin{equation}
MD_k = \sum_{j=1}^M{\left(\frac{m_{jk}-m_{jk^*}}{s_j}\right)^2}
\end{equation}
where $m_{jk^*}$ are the coordinates for
star $k^*$. The Mahalanobis distances from 51 Peg for these stars
and their ranks are listed in Table 7, and
a histogram of the Mahalanobis distances is found in Figure \ref{fig_Maha}.

Principal Components Analysis (PCA) is a statistical method for
partitioning total variability in a sample into linear combinations
that are orthogonal to each other
\citep{1999MNRAS.303..284R}.
The
first principal component is generated to maximize the variation of
that linear combination; the second principal component is chosen
orthogonal to the first and with maximum variation conditional on
the orthogonality. For these data the four moments, $m_{1k}, m_{2k}, m_{3k}, m_{4k}$,
for each star $k$ again form the basis for the calculations.

\begin{figure}
\includegraphics[scale=1.0]{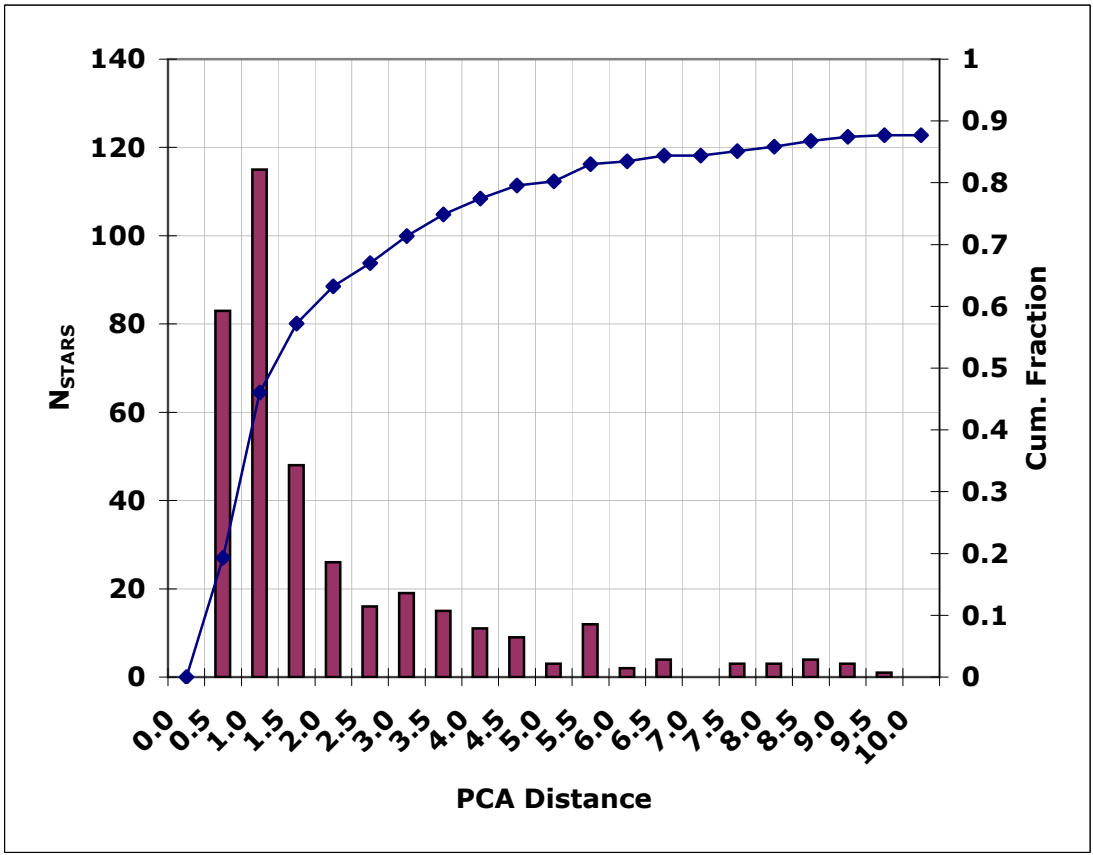}
\caption{\label{fig_PCA} Histogram of Principal Component Analysis distances from HD217014 for potential calibrator stars.}
\end{figure}

The PCA is carried out on the correlation matrix so at this stage we
do not need to adjust for the variances. The principal component for
51 Peg for the basis for the distance calculation of each star from
51 Peg. Specifically, let $p_{jk}$ be the principal components for
star $k$ and $p_{jk^*}$ the principal component for reference star
$k^*$. Then the distance star $k$ is from the reference star is,
\begin{equation}
PD_k = \sum_{j=1}^M{\left(\frac{p_{jk}-p_{jk^*}}{sp_j}\right)^2}
\end{equation}
where $sp_j$ is the standard deviation for the $j$th principal
component. The PCA distances and their ranks are given in Table 7, and
a histogram of the PCA distances is found in Figure \ref{fig_PCA}.

\begin{figure}
\includegraphics[scale=1.0]{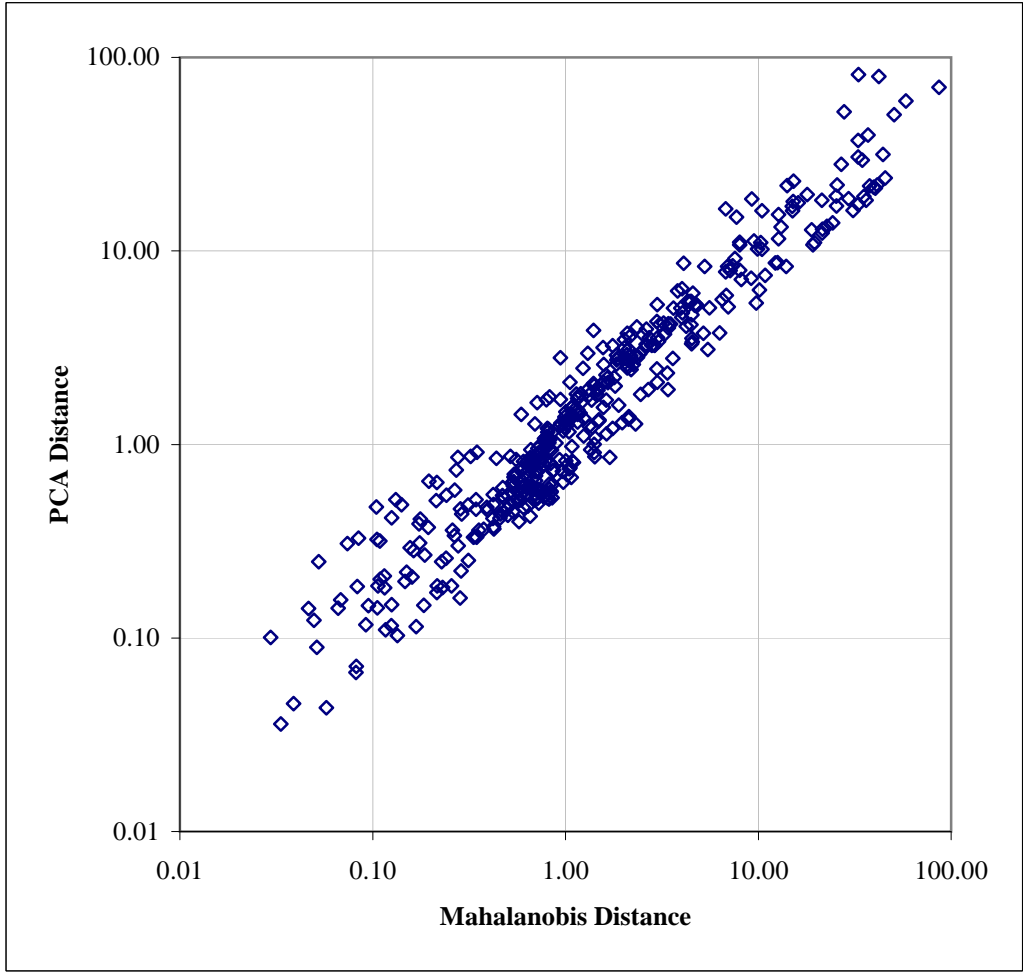}
\caption{\label{fig_PCAMaha} Principal Component Analysis versus Mahalanobis distances from HD217014 for potential calibrator stars.}
\end{figure}

The choice between Mahalanobis distance
and the PCA-based distance will be based on
convenience. The Mahalanobis distance has the advantage that it is
more immediately linked to the four moments of the distribution.
There is a very close correlation between the Mahalanobis distance
and the PCA-based distance as seen in Figure \ref{fig_PCAMaha}.
The PCA based approach has the advantage that the contribution of
each of the principal components can be calculated - e.g., for the
data considered here the first principal component accounts for 53 \% of the total
variability.

We find that the first of the known PTI binaries to appear among the ranked
potential calibrators of Table \ref{tab_BinaryRank} is HD 178449 - not unsurprising
given its large $\Delta K$ value.  The large brightness ratio makes the
object appear more like a calibrator than the other binary stars, and at
distances of $MD_{\textrm{\tiny HD178449}}=1.20$, $PD_{\textrm{\tiny HD178449}}=1.83$.
Consideration of the HD 178449 Maha and PCA ranks and an examination of
Figures 4 and 5 provides insight into the distributions for
both Maha and PCA distances.  Given that the sources were pre-selected to {\it likely}
be suitable as calibration sources, it is reasonable to expect that distributions
should peak at values corresponding to suitable calibrators.  If we select all
stars that fall towards the smaller values of those distribution, we can have
some confidence that we have statistically selected the stars that have been
demonstrated in this dataset to be suitable for use as calibrators.
As such, we established our cutoff at $MD, PD < 1.0$.  The resulting
calibrator list is available online at the MSC web site\footnote{http://msc.caltech.edu}.

It is important to note that the `rejection' of a star as a suitable calibrator
from this analysis does not necessarily indicate it is a heretofore unresolved
binary - merely that, within this particular dataset, the data indicate it is
unsuitable for use as a calibrator.  This {\it could} be due to some previously
undetected binarity, but could also be merely due to poor quality of data for that
particular object in the dataset.  A key indicator of this possibility is the fact that the
stars in Table \ref{tab_BinaryRank} that are listed as `acceptable' have, on average, a
larger expected angular size (0.449 mas) than those in the `reject' category (0.301 mas).
The smaller stars are, on average, more distant, and hence, are also on average dimmer - resulting in
lower signal-to-noise in the PTI system.  Such objects are more susceptible to poor seeing and/or weather conditions,
and are ultimately more likely to be rejected due to poor data quality that is unrelated to
actual system binarity.  For the purposes of this investigation, rejection of such sources is
acceptable in exchange for attempting to eliminate false positives for acceptable calibrators.

\begin{figure}
\begin{center}
\includegraphics[scale=0.60,angle=90]{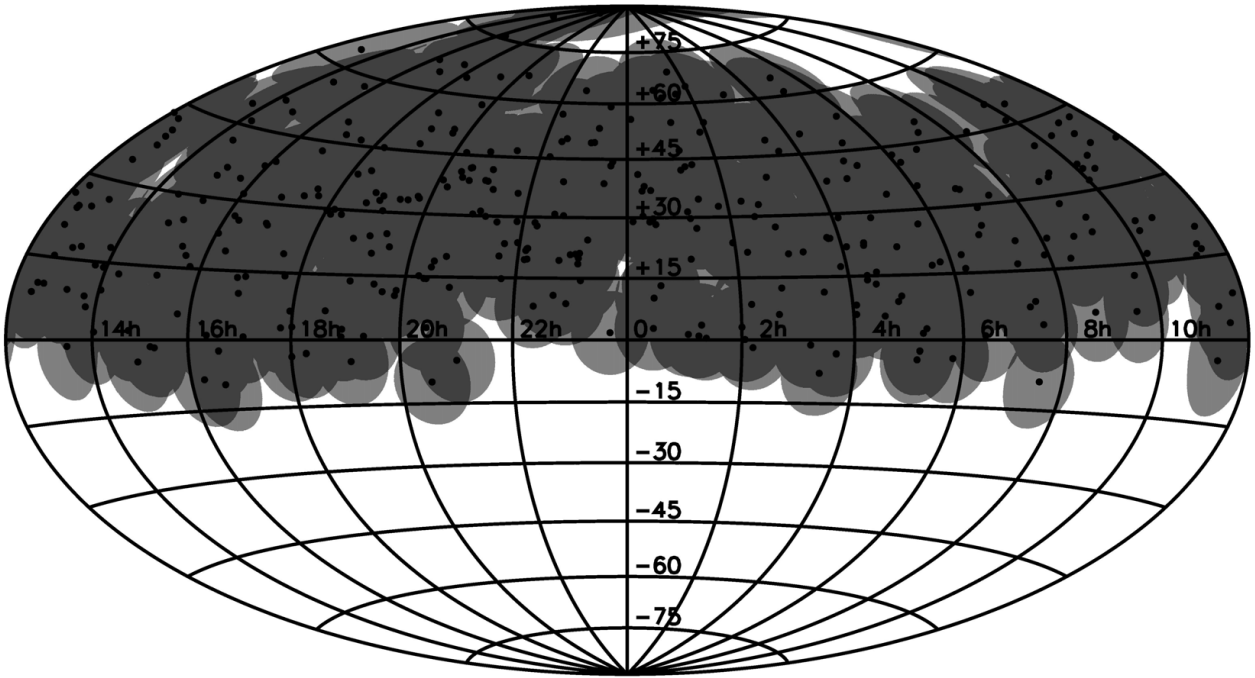}
\end{center}
\caption{\label{fig_ATOFF} Equatorial sky coverage map of statistically vetted calibrators displayed
in an equal area
Aitoff projection.  Light grey circles represent $10^o$ radii cones projected onto the sky,
representing regions of
appropriate calibrator proximity to astrophysical targets; darker grey regions represent
areas with two or more calibrators within $10^o$.  For PTI's nominal sky accessibility
of $5^o < \delta < 55^o$, these calibrators represent $\gtrsim 96$\% sky coverage.}
\end{figure}



\begin{deluxetable}{rrrrrrl}
\tablecolumns{7}
\tabletypesize{\scriptsize}
\tablewidth{0pc}
\tablecaption{Mahalanobis distance rank and principal components analysis rank of the potential
calibrator stars.\label{tab_BinaryRank}}
\tablehead{
\colhead{HD} & \colhead{Mahalanobis} & \colhead{Mahalanobis} & \colhead{PCA}    & \colhead{PCA}      & \colhead{Geometric}& \colhead{Binary} \\
\colhead{}   & \colhead{Rank}        & \colhead{Distance}    & \colhead{Rank}   & \colhead{Distance} & \colhead{Non-detection\tablenotemark{a}}& \colhead{Reference}
 } \startdata

\multicolumn{7}{l}{\bf Suitable as calibrators ($MD$ {\it and} $PD<1.0$):} \\
        71 &         94 &       0.51 &         82 &       0.49 &               0.036  &            \\

      1404 &         63 &       0.28 &         74 &       0.47 &                   -     &            \\

      1671 &        134 &       0.66 &        190 &       0.94 &                   -     &            \\

      2344 &         30 &       0.13 &         64 &       0.42 &               0.010  &            \\

      2454 &        172 &       0.79 &        203 &       1.04 &               0.000  &            \\

      2628 &        135 &       0.66 &        173 &       0.83 &                   -     &            \\

      2942 &         66 &       0.31 &         79 &       0.49 &               0.000  &            \\
\enddata
\tablenotetext{a}{Probability of a binary system non-detection due to lack of total nights on the sky, as discussed in  \S \ref{sec_UndetectedBinaries}.}
\end{deluxetable}



\section{Estimation of the Undetected Binary Rate}\label{sec_UndetectedBinaries}

One primary concern with the selection of potential calibration sources is choosing objects that
are previously undetected binaries.  For this reason, it is standard operating procedure
to observe no less than 2 calibration sources in conjunction with astrophysical targets,
allowing some means of cross-calibration between the potential calibrators.
Although we cannot know for certain which
of our previously unobserved calibration sources might have a heretofore undetected companion, we may attempt
to characterize the likelihood with which such an undetected companion would escape
detection once such a binary star system were to be placed on the observing queue.

We generated a synthetic, random sample of 4096 binary stars, starting with selecting
an object with a random spectral type consistent with our potential calibrator list, which is within
the range of B8 to K1.  Luminosity class for these objects was statically assigned to be either III or V, to test
the sensitivity of binary detection to primary star luminosity class.  Absolute V, K magnitudes, radii and masses for these primary
stars were then drawn from the standard MK type values found in \citet{bes88} and \citet{cox00}; apparent magnitude was
randomly assigned a value between $m_K = 3.5$ to $5.0$, which is consistent with the
apparent magnitudes of PTI calibrators.  Secondary stars were then randomly generated for
each of these primaries: spectral type was selected to be equal to or later than the primary star,
down to a class of M9\footnote{Implicit in this
condition is a `flat' mass ratio $q=M_1/M_2$ for binary stars, a topic which is discussed extensively
in \citet{duq91} and more recently in \citet{gol03}, which appears to be reasonable for this experiment.},
and for giant star primaries, luminosity class of the secondary objects was randomly assigned either III or V.
Absolute V, K magnitudes, radii and masses for these secondary
stars were then estimated also using the same references.  Orbital separation $a$ was then randomly assigned
to be between 2 and 50 Roche diameters; distance of the binary star pair was established from
the distance modulus, with an apparent angular separation resulting from the distance and true separation
of the primary and secondary.  Periods were established from the physical separation and the masses.
Orbital parameters, such as eccentricity $e$, inclination $i$, and longitude of the ascending node $\Omega$ were
randomly assigned.  Angular sizes of each component were estimated using the $V-K$ technique described in \citet{van99b};
given the $m_K = 3.5 - 5.0$ prior, the primaries were all well within the PTI calibrator selection criterion of $\theta<1.0$ mas.
Additionally, this sample well represents a potential sample of overlooked binary stars, with $a<0.1$ arcsec,
regardless of luminosity class.

Our overall goal was to test this subsample for PTI's ability
to detect deviations from what was expected to be normal calibrator object behavior.
The expected squared visibility in a narrow passband for a binary system is given by
\begin{equation}\label{eqn_binaries}
V^2(\lambda) = { V_1^2 + V_2^2r^2 +2 V_1 V_2 r \cos [ (2 \pi / \lambda){\bf B \cdot s}]\over (1+r)^2}
\end{equation}
where $V_1$ and $V_2$ are the visibility moduli for the two stars
alone as given by Equation \ref{eqn_UDdisk}, $r$ is the apparent
brightness ratio between the primary and secondary, $B$ is the
projected baseline vector at the system sky position, and $s$ is the
primary-secondary angular separation vector on the plane of the sky
\citep{hum95, bod99a}. $V_1$ and $V_2$ were estimated from the angular
sizes estimates. A
conservative detection limit would be a $3-\sigma$ threshold of
$\Delta V^2>0.06$, roughly three times PTI's limiting $V^2$
precision of $\sim 0.015-0.020$, where $\Delta V^2$ is the
difference between the expected single star $V^2$ as described in
Equation \ref{eqn_UDdisk} and the binary star $V^2$ from Equation
\ref{eqn_binaries}.  Objects with binary $V^2$ excursions less than
PTI's limiting $V^2$ precision would effectively be single stars as
far as PTI is concerned, and were excluded from consideration.

Figuring prominently into this characterization of PTI's ability to detect these synthetic binary stars is the
geometric orientation of the target systems relative to the instrument's baseline in use, contained in the
${\bf B \cdot s}$ term of Equation \ref{eqn_binaries}. If PTI's baseline in use is orthogonal to the
primary-secondary separation vector, the departure from point-source source visibility
will potentially be undetectable, depending upon the brightness ratio.
Since the relative geometry between binary system and projected baseline varies with orbital phase,
these 4096 systems were then examined over the course of 20
random epochs between the years covered by this study, 1998-2005,
using PTI configurations consistent with the available baselines.  As the number of nights increases,
the likelihood that a binary system would remain undetected is given in Table \ref{tab_BinaryDetnFitting}.

Two categories of binary stars were problematic for PTI: First, stars with detectable $V^2$ excursions ($\Delta V^2_{max}\geq 0.06$)
but in poor geometric orientation, due to orbital phase or apparent sky separation relative to baseline used,
are found to escape initial detection by the array.  However, after a few ($\sim 5$) nights of observing,
virtually all of these objects exhibited a detectable excursion and would be detected as binaries.
Second, and more problematic, are the stars with $V^2$ excursions below the $3-\sigma$ threshold
of obvious detectability, but above PTI's limiting $V^2$
precision of $\sim 0.015-0.020$.  These objects would remain always unconfirmed as binaries, yet
their varying $V^2$ response could affect system calibration at the 2-4\% level.



\begin{deluxetable}{ccrrrrrrrrr}
\rotate
\tablecolumns{11}
\tabletypesize{\scriptsize}
\tablewidth{0pc}
\tablecaption{Percentage of undetected binaries from the synthetic binary star subsample, by number of
nights observed per baseline and baselines used, as discussed in \S \ref{sec_UndetectedBinaries}.  Average value
for the brightness ratio is given for those binary systems expected to be detected and undetected by PTI.\label{tab_BinaryDetnFitting}}
\tablehead{
\colhead{}            & \colhead{Primary}    & \multicolumn{7}{c}{\underline{Nights Observed}} & \colhead{Undetected}& \colhead{Detected}\\
\colhead{Baseline(s)} & \colhead{lum. class} & \colhead{1 (\%)} & \colhead{2 (\%)} & \colhead{3 (\%)} & \colhead{4 (\%)} & \colhead{5 (\%)} & \colhead{6 (\%)} & \colhead{20 (\%)}& \colhead{Brightness Ratio} & \colhead{Brt. Ratio}
 } \startdata


  NS,NW,SW &          V &  $ 30.0 \pm 0.4 $  &  $ 13.6 \pm 0.4 $  &  $ 8.9 \pm 0.2 $  &  $ 7.2 \pm 0.1 $  &  $ 6.4 \pm 0.1 $  &  $ 5.9 \pm 0.0 $  &  $ 5.0 \pm 0.2 $  &  $ 0.00967 \pm 0.00035 $  &  $ 0.312 \pm 0.004 $  \\
  NS,NW,SW &        III &  $ 34.6 \pm 0.7 $  &  $ 21.2 \pm 0.7 $  &  $ 16.6 \pm 0.6 $  &  $ 14.8 \pm 0.7 $  &  $ 13.9 \pm 0.6 $  &  $ 13.3 \pm 0.7 $  &  $ 12.0 \pm 0.6 $  &  $ 0.00543 \pm 0.00004 $  &  $ 0.289 \pm 0.003 $  \\
        NS &          V &  $ 29.6 \pm 0.5 $  &  $ 13.3 \pm 0.5 $  &  $ 8.6 \pm 0.4 $  &  $ 6.8 \pm 0.2 $  &  $ 6.0 \pm 0.2 $  &  $ 5.6 \pm 0.3 $  &  $ 4.7 \pm 0.3 $  &  $ 0.01012 \pm 0.00114 $  &  $ 0.311 \pm 0.004 $  \\
        NS &        III &  $ 34.6 \pm 0.7 $  &  $ 21.5 \pm 0.3 $  &  $ 16.6 \pm 0.3 $  &  $ 14.7 \pm 0.2 $  &  $ 13.7 \pm 0.2 $  &  $ 13.1 \pm 0.3 $  &  $ 11.9 \pm 0.3 $  &  $ 0.00545 \pm 0.00012 $  &  $ 0.296 \pm 0.008 $  \\

\enddata
\end{deluxetable}



For example, in a single night of observing with the NS baseline has approximately $\sim 30\%$ of the
binary systems with a main sequence primary are undetected.  However, this proportion drops rapidly to $\sim 5.5\%$ after
only 5 nights of observing, and to
$\sim 5\%$ after 20 nights of observing, as it does for using the three PTI baselines in sequence.
For binary systems with giant primaries, the limiting proportion of undetected binaries is higher,
at $\sim 12\%$ after 20 nights.  This larger value is consistent with a greater number of binary systems with a large
brightness ratio, specifically those systems with a main sequence secondary star.  This non-zero proportion of
undetected binary systems is compelling motivation to employ multiple calibrators
for any study with PTI, or any other interferometer.
As we shall see in \S \ref{sec_StatTests}, we have many stars that show PTI $V^2$ data that
are consistent with the point-response
of a single, rather than binary, star system.  The lingering uncertainty of undetected
binarity for minimally observed stars
can be drawn
from the percentages given in Table \ref{tab_BinaryDetnFitting}.

\section{Discussion}

Some of the objects categorized as `acceptable' calibrators due to their PTI $V^2$ are still found
to have references indicating possible binary nature or other astrophysical effects which may cause them
to be less-than suitable as calibrators; these objects are discussed below in \S
\ref{sec_acceptedCalibrators}.
Interferometer observations of hierarchical
systems have the potential to accidentally observe the wrong star and so appear to have a
variation in angular size.   Variations in metallicity will mean a potential calibrator
won't fit SED templates well
for predicting size.   Stars with extreme rotational velocity will potentially exhibit
angular size variations depending upon interferometer baseline utilized, and potentially
``thrown off'' material (as seen with the Be star phenomena) will also affect visibility
measurements if the disk is detected.
Additionally,
evidence in the literature was uncovered for rejected calibrator candidates
and is discussed in \S \ref{sec_rejectedCalibrators}.

\subsection{Objects thought to be acceptable calibrators}\label{sec_acceptedCalibrators}
HD 905 is listed as a periodic variable star \citep{koen02}.

HD 1279 shows P Cygni type profiles in UV spectra taken with IUE \citep{snow94}.

HD 2758 is listed as a binary in the Tycho-2 catalog \citep{fabr02} with a companion at position angle 69.6
degrees and separation of 0\farcs44.

HD 3268 is listed in \cite{cayr97} as metal poor with [Fe/H] = -0.23.

HD 15335 is listed in \citet{1994AJ....108.2338H} as an 0\farcs020 astrometric binary,
detected via photographic plates obtained with the 61 cm Sproul refractor, although the comments cite it as ``astrom.,probable'',
and no secondary component magnitude is listed.  Follow-up work by \citet{2005ApJ...622.1102F} report a null result not only on detecting a
stellar companion, but also down to planetary mass.

Both HD 19994 and HD 90508 are indicated to be binary stars in \citet{1994AJ....107..306H}.  HD 19994 is notable in that it has a
Jupiter-mass planet \citep{2004A&A...415..391M}, and the F8V star's stellar companion is an M3V dwarf a few arcseconds away, which makes it undetectable
by PTI from a brightness ratio standpoint.  Similarly, HD 90508 is a F9V star with a M3V companion.

HD 24357 is listed in \cite{cayr97} as metal rich with [Fe/H] = 0.30.

HD 27524 is listed in \cite{cayr97} as metal poor with [Fe/H] = -0.46.

HD 28024 is listed in the Washington Double
Star Catalog as a visual double with 104\farcs3 separation
and a delta magnitude of 8.23 \citep{worl96}.  It has a GCVS
period of 0.148 days \citep{samus04} and in addition, it is cited as a delta Scuti
variable \citep{rodr00}. It also has high rotational velocity (see below).

HD 28677 is listed in the GCVS as a possible variable star, however,
no period or variability type is identified and this designation may
be in doubt \citep{samus04}. We also note that this star is listed in
a catalogue of binaries detected using lunar occultation and has a
separation from its companion of 16.2 $\pm$ 5.2 mas, at a position
angle of 205.9 degrees, with a delta magnitude of 0.0 $\pm$ 1.62
magnitudes at 547 nm \citep{mason95}. There is no evidence for
binarity in the PTI data, despite the 9 nights of available
calibrated PTI data (7 on the NS baseline, 2 on the NW baseline).

HD 28704 is listed in the Washington Double Star Catalog as being a hierarchical quadruple system with
separations from the primary ranging from 76\farcs4 to 113\farcs7 and delta magnitudes
with respect to the primary ranging from 0.71 to 7.3 \citep{worl96}.

HD 31662 is listed in the Washington Double Star Catalog as being a
visual double with separation 5\farcs4 and a delta magnitude of 5.47
\citep{worl96}.  There is no evidence for binarity in the PTI data.

HD 33167 has a large range of metallicities in the literature, from [Fe/H] = 0.09 \citep{cayr97} to
[Fe/H] = -0.36 to -0.37 (\citet{ibuk02} and \citet{mars95}, respectively).

HD 38558 is noted by \cite{adel01} to have variability of 0.01 magnitudes in the Hipparcos data.

HD 43042 is listed in the Washington Double Star Catalog as being a hierarchical quadruple system
with separations from the primary ranging from 32\farcs0 to 91\farcs2 and delta magnitudes with
respect to the primary ranging from 4.8 to 6.0 \citep{worl96}.

HD 51530 is listed in the Washington Double Star Catalog as being a
hierarchical triple system with separations from the primary of
0\farcs5 and 28\farcs2 and a delta magnitude of the larger separated
component of 6.0 magnitudes \citep{worl96}.  There is no evidence for
binarity in the PTI data.

HD 58551 is noted to be extremely metal poor, with [Fe/H] = -0.6 \citep{bart84,cayr97}.

HD 58715 is noted as a double in the Hipparcos Input Catalogue with unknown separation and
magnitude difference \citep{turon93}. It is listed as an eruptive
Gamma Cas variable by \citep{samus04}
and has high rotational velocity (see below).

HD 58946 is listed as a hierarchical quadruple system spread over 5 arcminutes.  The closest
companion to the primary is separated by 3\farcs4 with a delta magnitude of 8.4 \citep{turon93}.
\citet{worl96} goes on to list the A component in the system as a spectroscopic binary.  However,
\citet{gall05} use the system as a constant RV source to demonstrate the stability of their
ELODIE instrument in searching for brown dwarfs and extrasolar planets. \citet{samus04} lists
it in the GCVS with unknown variability type.  And finally \citet{adel01} shows it to have 0.01 magnitudes
of variability, as measured by Hipparcos.

HD 71148 is listed in the GCVS \citep{samus04} as a previously misidentified variable star.  It
is metal poor with [Fe/H] = -0.15 \citep{ibuk02}.

HD 75332 has been studied extensively in radial velocity surveys searching for extrasolar
planets, however, no evidence for any are found to date \citep{2005ApJ...622.1102F}.

HD 75732 is a visual double with a separation of 85\farcs0 and a delta magnitude of 7.2.
It has been found to have planets \citep{2005ApJ...622.1102F}, which are well below the
sensitivity threshold of PTI to detect.

HD 89125 is a visual double with a separation of 7\farcs4 \citep{turon93}. Additionally, it is
seen to have low-metallicity ranging from [Fe/H] = -0.19 to -0.39 \citep{cayr97,mars88}.

HD 97334 is a hierarchical quadruple system with separations from the primary ranging from 88\farcs2
to 128\farcs4 and magnitude differences from 1.5 to 6.0 \citep{worl96}.

HD 97633 is flagged as a variable star by \citet{samus04} with the variable type and period unidentified.
Further, it is extremely metal-rich with [Fe/H] = 0.40 \citep{cayr97}.

HD 103095 is flagged as a suspected varaible star by \citet{samus04} with the variable type and
period unidentified.  It has also been studied by \citet{lath02} for binarity and was found to
have radial velocity measurements which varied at the 2-3 km s$^{-1}$ level, with no indication
of binarity or an orbital solution.

HD 119550 has also been studied by \citet{lath02} for binarity and was found to have radial
velocity measurements which varied at the 1 km s$^{-1}$ level, with no indication of
binarity or an orbital solution.

HD 128167 is flagged as a suspected variable star by \citet{samus04} with the variable type,
based on a reference, potentially a sigma bootes variable with an unidentified period.  It is
identified in the Washington Double Star catalog as a member of a hierarchical triple
system with separations from the primary of 248\farcs0 and
237\farcs0 and 5.34 and 6.80 magnitudes respectively \citep{worl96}. Finally, it displays
a wide range of low metallicities, from [Fe/H] = -0.18 to -0.60 \citep{cayr97}.

Both HD 132254 and HD 134083 are flagged by \citet{nor04} as spectroscopic binaries,
although no other supporting evidence in the literature could be found
for this conclusion for either star.
Indeed, in \citet{2005A&A...443..337G}, HD 134083 is cited as having a constant radial velocity at the $\sim 50$ m/s level,
which would preclude any binary detectable by PTI.  HD 132254
was well-behaved enough for \citet{2004AN....325..439K} to list the object as a candidate RV standard.

HD 136118 has been studied extensively with radial velocity techniques and is know to have
a planet \citep{santos03, santos04}, which are well-below
the sensitivity threshold of PTI to detect.

HD 141187 is studied in a catalog of astrometric binaries, identified using Hipparcos and Tycho
data, to have nonlinear proper motion \citep{maka05}.  It is also seen to have high rotational
velocity (see below).

HD 154345 is listed in \citet{samus04} as a suspected variable star
in the GCVS, however no period of variability type is listed.

HD 166205 is flagged in \citet{mason99} as a suspected binary based on ground-based and
Hipparcos data, however the binarity was not able to be confirmed, so it is listed in
their Table 7 as a "problem star". Finally, it is flagged in \cite{adel01} as a variable
star in the Hipparcos survey with 0.01 magnitudes of variability.

HD 171834 is flagged as a binary with 0\farcs1 arcsecond separation and no notation of the
secondary's magnitude in \citet{worl96}.  It is also seen to have a low metallicity compared
to solar of [Fe/H] = -0.42 \citep{cayr97}.

HD 182488 has been studied extensively using the radial velocity technique and is found to
have no detectable planet or stellar companions \citet{2005ApJ...622.1102F}.

HD 186568 is a member of a hierarchical triple with separations of 15\farcs4 and 33\farcs9
and magnitude differences of 7.75 and 5.44 respectively \citep{worl96}.

HD 187691 is a member of a hierarchical quintuple system with
separations from the primary ranging from 20\farcs5 to 89\farcs7 and
magnitude differences of 7.69 to 8.59 \cite{worl96}. It was used by
\cite{mazeh96} as an RV standard in the study of orbits of three
spectroscopic binary systems.  \citet{2005ApJ...622.1102F} have
studied it extensively in their radial velocity searches for planets
and find no evidence for a companion.  And finally, it is flagged in
the GCVS as a rotating variable star with no period listed
\citep{samus04}.

HD 187923 is a member of a visual double system with a separation from the secondary
of 93\farcs9 and a magnitude difference of 5.57 \citep{worl96}.  It has been studied
extensively using radial velocity techniques and is found to have no companions
\citet{2005ApJ...622.1102F}.  It is flagged in the GCVS as variable, but no variability
type or period is identified \citep{samus04}.

HD 195564 is a member of a hierarchical triple system with separations of 4\farcs6 and
100\farcs2 and magnitude differences of 5.5 and 4.2 respectively \citep{worl96}. It has
been studied extensively using radial velocity techniques and is found to have no
companions \citet{2005ApJ...622.1102F}.

HD 197076 is a member of a hierarchical triple system with
separations of 98\farcs8 and 125\farcs0 and magnitude differences of
5.45 and 6.95 respectively \citep{worl96}. It has been studied
extensively using radial velocity techniques and is found to have no
companions \citet{2005ApJ...622.1102F}.

HD 198478 is a member of a visual double system with a separation of 20\farcs5 and a
magnitude difference of 5.32 \citep{worl96}.  It is flagged in the GCVS as being
a pulsating variable star \citep{samus04} and has an identified period of
0.20472 days \citep{koen02}.

HD 210855 is a member of hierarchical triple system with separations of 79\farcs6 and
56\farcs0 and magnitude differences of 5.26 and 8.46 respectively \citep{worl96}.

HD 214680 is a member of a visual double system with a separation of 60\farcs4 and
a magnitude difference of 5.14 \citep{worl96}. Further, it is flagged in the GCVS
as a beta Cephei variable star \citep{samus04}.

HD 218396 is flagged in the GCVS as a rotating ellipsoidal variable \citep{samus04} and
as a probable lambda Bootis variable by \cite{gerb03}.

HD 219080 is flagged as a new periodic variable star by \cite{koen02}.

HD 222439 is a member of a hierarchical triple system with separations of 46\farcs6 and
103\farcs2 and magnitude differences of 7.16 for both companions \citep{worl96}. Further,
it is flagged in the GCVS as a variable star of unknown type \citep{samus04}.

In addition, the following stars have rotational velocities (v sin i) between 100 and 200 km/s: HD 7804, HD 18411,
HD 20150, HD 26605, HD 27397, HD 27946, HD 28677, HD 30739, HD 32301, HD 32630, HD 37147, HD 50019, HD 56537,
HD 60111, HD 67006, HD 79439, HD 83287, HD 85795, HD 87696, HD 92825, HD 102124, HD 107904, HD 108765,
HD 110411, HD 111604, HD 116831, HD 122408, HD 125161, HD 125162, HD 132052, HD 141003, HD 143894,
HD 147547, HD 152614, HD 161868, HD 166014, HD 166205, HD 168914, HD 169702, HD 177196, HD 177756,
HD 192425, HD 204153, HD 206043, HD 210129, HD 213558, HD 214734, HD 214923, HD 216735, HD 222439.
The following stars have rotational velocities higher than 200 km/s: HD 11946, HD 17573, HD 20418,
HD 21686, HD 23630, HD 23862, HD 28024, HD 30823, HD 58715, HD 73262, HD 93702, HD 98058,
HD 130109.  These data were taken mainly from Vizier catalogs associated with the following
papers: \cite{gleb00}, \cite{royer02} and \cite{deme02}.
Rapidly rotating stars have an additional uncertainty in their predicted angular size due
to rotational distortion, which is on the order of $\sim 10-15$\% \citep{2006ApJ...637..494V}.
However, if this was a significant factor, the statistical vetting of \S \ref{sec_StatTests}
would have rejected these objects as valid calibrators.

\subsection{Objects rejected as acceptable calibrators}\label{sec_rejectedCalibrators}

HD 8357 is flagged by \citet{nor04} as a spectroscopic binary, with a mass ratio of $q=0.841 \pm 0.009$.

HD 8799 was found to be listed as a visual (but not physical) binary of unspecified separation in \citet{tok02}.

HD 13480 is flagged by \citet{nor04} as a spectroscopic binary, with a mass ratio of $q=1.062 \pm 0.010$.

HD 16234 is flagged by \citet{nor04} as a spectroscopic binary, with a mass ratio of $q=1.039 \pm 0.041$.

HD 18012 was examined by \citet{1987AJ.....94.1318L} using speckle interferometry at the 4m Mayall telescope, without any detection of a secondary companion.

HD 27483 is flagged by \citet{nor04} as a spectroscopic binary, with a mass ratio of $q=0.937 \pm 0.011$.

HD 27901 was examined by \citet{1993AJ....105..220M} using speckle interferometry, without any detection of a secondary companion.

HD 37594 is found to have a range of spectral classifications from A5mF2 \citep{bert67} to F0Vp \citep{abt95}
as found by \cite{skiff05} indicating that the model fits to these data may be problematic.

HD 43587 is noted by \citet{2002ApJS..141..503N} as having a radial velocity RMS $>$ 0.1 km/s, and \citet{2005AJ....129.2420M} cite the object
as a possible astrometric binary from Hipparcos proper motion discrepancies,
with the factor $Q_0$ as discussed in that study, related to the lower bound of a possible secondary companion's
mass, having a value of $\log Q_0$=-1.6 yr$^{-1}$.

HD 75137 is listed by \cite{turon93} to have a companion separated by 12\farcs4 with a magnitude
difference of 7.5.  \cite{worl96} state that the primary in the system is a spectroscopic binary with
an orbital period of 8.2 days.  It also has high rotational velocity (see below).

HD 79969 is noted by \citet{2004AJ....127.1727H}, using speckle
interferometry, to have a secondary companion.  The orbital axis is 0\farcs66 $\pm$ 0\farcs007 with a period of 34.2 years \cite{henry93}.

HD 118232 is flagged as a suspected variable star by \cite{samus04}
with the variable type and period unidentified.  A visual inspection
of the PTI visibility data for the 9 nights this object was observed
shows 7 that are qualitatively identical to the associated
calibrators, and 2 that are marginally offset $\Delta V^2 \approx
0.05$ from those calibrators.

HD 120510 is flagged by \citet{nor04} as a spectroscopic binary,
with a mass ratio of $q=1.000 \pm 0.009$.

HD 122676 is noted by \citet{2002ApJS..141..503N} as having a radial velocity RMS $>$ 0.1 km/s.

HD 123999 is flagged by \citet{nor04} as a spectroscopic binary, with a mass ratio of $q=0.963 \pm 0.010$.

HD 141187 is noted by \citet{2005AJ....129.2420M} as a possible astrometric binary from Hipparcos proper motion discrepancies,
with $\log Q_0$=-1.1 yr$^{-1}$.

HD 149630 is noted by \citet{2005AJ....129.2420M} as a possible astrometric binary from Hipparcos proper motion discrepancies,
with $\log Q_0$=-1.0 yr$^{-1}$.

HD 152308 is noted in \citet{1991bsc..book.....H} as having ``suspected variable radial velocity''.
It appears in the GCVS catalog as a rotating variable star \cite{samus04} and has an identified
period of 0.9366 days \cite{rens01}.

HD 157935 has a radial velocity of -51.9 km s$^{-1}$ quoted in \citet{1995A&AS..114..269D} (no error given, but a radial velocity quality
of `B' on a scale of `A' to `E', where `A' is best), but a value of $-54.7 \pm 1.6$ km s$^{-1}$  listed in \citet{nor04} - possibly
an indication of a variable radial velocity.

HD 181655 is noted by \citet{2002ApJS..141..503N} as having a stable radial velocity with RMS $<$ 100 m/s.

HD 195050 is noted by \citet{2005AJ....129.2420M} as a possible astrometric binary from Hipparcos proper motion discrepancies,
with $\log Q_0$=-1.1 yr$^{-1}$. It is also seen to have high rotational velocity (see below).

HD 120048, HD 120509, HD 136643, HD 158063, HD 186547, and HD 234677 are all listed in the radial velocity study of
\citet{2005A&A...430..165F} as having ``no evidence for radial-velocity variations''.

HD 216538 is flagged by \cite{samus04} as a long-period pulsating B star.

HD 216831 is a member of a visual double system with a separation of
49\farcs6 and a magnitude difference of 3.76 \cite{worl96}. Further,
it was examined by \citet{mca87} using speckle interferometry at the
3.6m CFHT, without any detection of a secondary companion. Finally,
it has anomalously high metallicity compared to solar of [Fe/H] =
0.40 \citep{cayr97}.

HD 223346 is flagged to have anomalously low metallicity compared to solar of
[Fe/H] = -0.42 \citep{ibuk02}.

HD 223421 is flagged to have anomalously low metallicity compared to solar
ranging from [Fe/H] = -0.20 to -0.35 (see for example \cite{ibuk02} and \cite{cayr97}).
It is flagged as a probable spectroscopic binary in a catalog of
overluminous F-type stars, though identification was problematic \citep{grif03}.

In addition, the following stars have rotational velocities (v sin i) between 100 and 200
km/s: HD 20677, HD 23850, HD 27322, HD 27901, HD 45542, HD 75137, HD 82621, HD 141187,
HD 141851, HD 144874, HD 178449, HD 195050, HD 204403. The following stars have rotational
velocities higher than 200 km/s:  HD 149630 and HD 184606.
These data were taken mainly from Vizier catalogs associated with the following
papers: \cite{gleb00}, \cite{royer02} and \cite{deme02}.

\section{Conclusion}

We have examined 8 years of PTI visibility data for candidate
single, point-like stellar calibrator sources.  To vet the
appropriate sources, we subjected the data to a rigorous statistical
evaluation, comparing the sources to a well-understood and
intensively studied standard PTI calibrator, HD217014.  Of the
candidates that satisfied the {\it a priori} selection criteria,
approximately 350 were found to be `well-behaved' in an empirical,
statistical sense, with $\sim 140$ being rejected.  These vetted
calibrator objects represent $\gtrsim 96$\% sky coverage for PTI
in the declination range $5^o < \delta < 55^o$ and
form a set of well-characterized calibrator anchors for future PTI
observations, archival studies, and observations at other
interferometric facilities.

This research has made use of NASA's Astrophysics Data System. This
research has made use of the SIMBAD and VizieR databases, operated
at CDS, Strasbourg, France.
Photometric data was obtained in part from the General Catalog of
Photometric Data \citep{1997A&AS..124..349M}.
This publication makes use of data
products from the Two Micron All Sky Survey, which is a joint
project of the University of Massachusetts and the Infrared
Processing and Analysis Center/California Institute of Technology,
funded by the National Aeronautics and Space Administration and the
National Science Foundation.

Science operations with PTI are conducted through the efforts of the
PTI Collaboration (http://msc.caltech.edu/missions/Palomar/), and we
acknowledge the invaluable contributions of our PTI colleagues.
Funding for PTI was provided to the Jet Propulsion Laboratory under
its TOPS (Towards Other Planetary Systems), ASEPS (Astronomical
Studies of Extrasolar Planetary Systems), and Origins programs and
from the JPL Director's Discretionary Fund. Portions of this work
were performed at the Jet Propulsion Laboratory, California
Institute of Technology under contract with the National Aeronautics
and Space Administration.

\end{document}